\documentclass[useAMS,usenatbib]{mn2e}

\usepackage{times}

\newcommand{\one}{HD\,189733}	
\newcommand{\two}{HD\,209458}	
\newcommand{\thetaone}{$0.3848 \pm 0.0055$}	
\newcommand{\thetatwo}{$0.2254 \pm 0.0072$}	
\newcommand{\thetaoneud}{$0.3600 \pm 0.0046$}	
\newcommand{\thetatwoud}{$0.2147 \pm 0.0066$}	

\newcommand{\fboloneorig}{$2.785 \pm 0.015$}		
\newcommand{\fboltwoorig}{$2.331 \pm 0.020$}		
\newcommand{\fbolone}{$2.785 \pm 0.058$}		
\newcommand{\fboltwo}{$2.331 \pm 0.051$}		
\newcommand{\luminone}{$0.328\pm0.011$}	
\newcommand{\lumintwo}{$1.788\pm0.147$}		

\newcommand{\rhoone}{$1.62 \pm 0.11$}	
\newcommand{\rhotwo}{$0.58 \pm 0.14$}	
\newcommand{\rhoonep}{$0.605 \pm 0.029$}	
\newcommand{\rhotwop}{$0.196 \pm 0.033$}	
\newcommand{\rhoonecgsp}{$0.802 \pm 0.038$~g~cm$^{-3}$}	
\newcommand{\rhotwocgsp}{$0.260 \pm 0.043$~g~cm$^{-3}$}	

\newcommand{\radonep}{$1.216 \pm 0.024$}	
\newcommand{\radtwop}{$1.451 \pm 0.074$}	
\newcommand{\radone}{$0.805 \pm 0.016$}	
\newcommand{\radtwo}{$1.203 \pm 0.061$}	

\newcommand{\teffone}{$4875 \pm 43$}	
\newcommand{\tefftwo}{$6092 \pm 103$}	

\newcommand{\loggone}{$4.56 \pm 0.03$}	
\newcommand{\loggtwo}{$4.28 \pm 0.10$}	
\newcommand{\loggonep}{$3.29 \pm 0.02$}	
\newcommand{\loggtwop}{$2.88 \pm 0.07$}	

\newcommand{\rsun}{R$_{\odot}$}			
\newcommand{\rjup}{R$_{\rm Jup}$}		
\newcommand{\rhojup}{$\rho_{\rm Jup}$}	
\newcommand{\rhosun}{$\rho_{\odot}$}	
\newcommand{\msun}{M$_{\odot}$}			
\newcommand{\lsun}{L$_{\odot}$}			
\newcommand{\teff}{$T_{\rm eff}$}			
\newcommand{\amlt}{$\alpha_{\rm MLT}$}			
\newcommand{\um}{$\mu$m}

\newcommand{\aster}{asteroseismology}

\newcommand{\phs}{\phantom{$-$}}	
\newcommand{\phn}{\phantom{0}}		

\def\apjl{{ApJ}}                
\def\apjs{{ApJS}}               
\def\aap{{A\&A}}                

\usepackage{epsfig}
\usepackage{graphicx}              
\usepackage{afterpage}
\usepackage{deluxetable}
\usepackage{natbib}                

\usepackage{times}
\usepackage{array}
\usepackage{color}
\usepackage{hyperref}
\definecolor{Blue}{rgb}{0.3,0.3,0.9}

\title[Fundamental Parameters of Exoplanet Host Stars]
{Stellar Diameters and Temperatures VI. \\High angular resolution measurements of the transiting exoplanet host stars HD~189733 and HD~209458 and implications for models of cool dwarfs}
\author[Boyajian, von Braun, et al.]
{\parbox{\textwidth}{Tabetha Boyajian$^{1}$\thanks{E-mail: tabetha.boyajian@yale.edu},
Kaspar von Braun$^{2,3,4}$,
Gregory A. Feiden$^{5}$,
Daniel Huber$^{6,7}$,
Sarbani Basu$^{1}$,
Pierre Demarque$^{1}$,
Debra A. Fischer$^{1}$,
Gail Schaefer$^{8}$,
Andrew W. Mann$^{9,10}$,
Timothy R. White$^{11}$,
Vicente Maestro$^{12}$,
John Brewer$^{1}$,
C. Brooke Lamell$^{1}$,
Federico Spada$^{13}$,
Mercedes L\'{o}pez-Morales$^{14}$,
Michael Ireland$^{15}$,
Chris Farrington$^{8}$,
Gerard T. van Belle$^{4}$,
Stephen R. Kane$^{16}$,
Jeremy Jones$^{17}$,
Theo A. ten Brummelaar$^{8}$,
David R. Ciardi$^{18}$, 
Harold A. McAlister$^{17}$,
Stephen Ridgway$^{19}$,
P. J. Goldfinger$^{8}$,
Nils H. Turner$^{8}$,
and Laszlo Sturmann$^{8}$
}
\vspace{0.3cm}\\
$^{1}$Yale University, New Haven, CT, USA\\
$^{2}$Max-Planck-Institute for Astronomy (MPIA), K\"{o}nigstuhl 17, 69117 Heidelberg, Germany\\
$^{3}$Mirasol Institute, Munich, Germany\\
$^{4}$Lowell Observatory, Flagstaff, USA\\
$^{5}$Department of Physics \& Astronomy, Uppsala University, Box 516, SE-751 20 Uppsala, Sweden\\
$^{6}$NASA Ames Research Center, Moffett Field, CA 94035, USA\\
$^{7}$SETI Institute, 189 Bernardo Avenue, Mountain View, CA 94043, USA\\
$^{8}$The CHARA Array, Mount Wilson Observatory, Mount Wilson, CA 91023, USA\\
$^{9}$Harlan J. Smith Fellow\\
$^{10}$Department of Astronomy, The University of Texas at Austin, Austin, TX 78712, USA\\
$^{11}$Institut f\"{u}r Astrophysik, Georg-August-Universit\"{a}t G\"{o}ttingen, Friedrich-Hund-Platz 1, 37077 G\"{o}ttingen, Germany\\
$^{12}$Sydney Institute for Astronomy, School of Physics, University of Sydney, NSW 2006, Australia\\
$^{13}$Leibniz-Institut f\"{u}r Astrophysik Potsdam (AIP), An der Sternwarte 16, 14482, Potsdam, Germany\\
$^{14}$Harvard-Smithsonian Center for Astrophysics, 60 Garden Street, Cambridge, MA 03128, USA\\
$^{15}$Research School of Astronomy \& Astrophysics, Australian National University, Canberra ACT 2611, Australia\\
$^{16}$Department of Physics and Astronomy, San Francisco State University, 1600 Holloway Ave., San Francisco, CA 94132, USA\\
$^{17}$Center for High Angular Resolution Astronomy and Department of Physics and Astronomy, Georgia State University, Atlanta, GA, USA\\
$^{18}$NASA Exoplanet Science Institute, California Institute of Technology, MC 100-22, Pasadena, CA 91125, USA\\
$^{19}$NOAO, Tucson, AZ, USA
}

\begin{document}

%

\maketitle

\label{firstpage}

%

\begin{abstract}

\noindent

We present direct radii measurements of the well-known transiting exoplanet host stars \one~and \two~using the CHARA Array interferometer.  We find the limb-darkened angular diameters to be $\theta_{\rm LD} =$~\thetaone~and \thetatwo~milliarcsec for \one~and \two, respectively.  \one~and \two~are currently the only two transiting exoplanet systems where detection of the respective planetary companion's orbital motion from high resolution spectroscopy has revealed absolute masses for both star and planet.  We use our new measurements together with the orbital information from radial velocity and photometric time series data, {\it Hipparcos} distances, and newly measured bolometric fluxes to determine the stellar effective temperatures ($T_{\rm eff} =$~\teffone, \tefftwo~K), stellar linear radii ($R_{\ast} =$~\radone, \radtwo~\rsun), mean stellar densities ($\rho_{\ast} =$~\rhoone, \rhotwo~\rhosun), planetary radii ($R_{\rm p} =$~\radonep, \radtwop~\rjup), and mean planetary densities ($\rho_{\rm p} =$~\rhoonep, \rhotwop~\rhojup) for \one b and \two b, respectively.  The stellar parameters for \two, a F9 dwarf, are consistent with indirect estimates derived from spectroscopic and evolutionary modeling. However, we find that models are unable to reproduce the observational results for the K2 dwarf, \one.  We show that, for stellar evolutionary models to match the observed stellar properties of \one, adjustments lowering the solar-calibrated mixing length parameter to \amlt~$= 1.34$\ need to be employed. 

\end{abstract}




\section{Introduction}\label{sec:introduction}

Exoplanet characterization relies heavily on our ability to accurately describe the host star properties, as the properties of a planet are only known as well as those of its host star. A common approach is to use stellar atmosphere and evolutionary models to determine stellar properties from observables like spectral features and/or photometric colors. However, the comparison between these indirect calculations and direct measurements of both single and binary star radii and temperatures have consistently produced a discrepancy: directly determined values tend to be $\sim 5$\% larger and $\sim 3$\% cooler than their corresponding values predicted by models (e.g. \citealt{tor10,boy12b}).  The source of this discrepancy is still unclear, but suggested explanations include stellar age, magnetic activity/starspots, close binary interactions, composition, convection, equation of state, mixing length theory, solar mixtures, or combinations of the above factors not being properly accounted for in the modelling processes. 

Alleviating this dependence on models by directly measuring host star properties is a golden ticket to unbiased and {\it absolute} system properties. While empirical determination of the stellar radius is rare, select cases do exist where the host star radius is measured with long-baseline optical interferometry (LBOI) or \aster.  In the case of the former, LBOI has resolved the transiting exoplanet hosts GJ\,436 \citep{von12}, 55\,Cnc \citep{von11b, van09, bai08}, and \one~\citep{bai07a}.  Of these, only recent improvements to instruments and increased sensitivities and techniques have enabled measurements to determine these stellar radii to better than 5\% precision (3.1\%~for GJ\,436~\citealt{von12}; and 0.6\%~for 55\,Cnc~\citealt{von11b}). For both transiting and non-transiting exoplanet hosts, the combination of the stellar angular size from LBOI, trigonometric parallax from Hipparcos, and bolometric flux via spectral energy distribution fitting allows for largely model-independent determination of stellar radii and effective temperatures (e.g. \citealt{von14}).  

The latter technique of using \aster~to measure radii of transiting exoplanet host stars has been shown to be a fruitful resource in recent years compared to LBOI.  The progress in this field is well described in \citet{hub13}, who present results from the NASA Kepler mission of 77 exoplanet host stars in the Kepler field that have radii and masses via \aster~with uncertainties of $\sigma (R_{\ast}) \sim 3$\% and $\sigma (M_{\ast}) \sim 7$\%.  Lastly we note that the detections of circumbinary planets, i.e. transiting planets in eclipsing binary systems, have enabled the extraction of stellar/planetary radii to high precision through a full photometric-dynamical model \citep[e.g., see][]{car11, doy11}. Unfortunately, although \citet{wel12} predict that 1\% of close binary stars should have planets in such a perfect viewing configuration, few systems are known or well characterized. 

This paper presents LBOI observations of two well-known, transiting, exoplanet host stars \one~($V$mag~$= 7.70$, K2~V; \citealt{gra03}) and \two~($V$mag~$= 7.65$, F9~V; \citealt{gra01}). We introduce our data in \S~\ref{sec:data}, and present the stellar and revised planetary properties in \S~\ref{sec:properties}.  In \S~\ref{sec:discussion}, we describe various model dependent stellar properties in comparison with this work. In \S~\ref{sec:modelcomp}, we discuss scenarios to reconcile the discrepant results of the data with models for the lower-mass host, \one.


\section{Data} \label{sec:data}


\subsection{Interferometric observations}\label{sec:observations}

Interferometric observations were performed with the CHARA Array, a long-baseline optical/infrared interferometer located at the historic Mount Wilson Observatory in California.  The CHARA Array consists of six 1-m diameter telescopes in a Y-configuration where the distances between telescopes, referred to as the baseline $B$, range from $\sim 30 - 330$~meters.  

The predicted angular sizes of \one~and \two~are on the order of a few tenths of a milli-arcsecond \citep[e.g., see][and the discussion below]{boy14a}. Thus, observations were conducted using the PAVO beam combiner with pairs of telescopes on the longest baseline configurations available in order to adequately resolve the stars.  The PAVO beam combiner operates in the $R$-band \citep{ire08}, and routinely measures precise stellar angular diameters well under a milli-arcsecond \citep{bai12, hub12, whi13, mae13}.

A log of the observations is shown in Table~\ref{tab:observations}.  In summary, observations of each object were bracketed in time with several calibrator stars. Initial query of suitable calibrators is based on the JMMC Stellar Diameters Catalog (JSDC; \citealt{bon06,bon11})\footnote{\href{http://www.jmmc.fr/catalogue\_jsdc.htm}{http://www.jmmc.fr/catalogue\_jsdc.htm}.}.  We selected calibrators based upon their physical attributes: no known duplicity, low projected rotational velocity\footnote{Stars become oblate if rotating near critical velocities.  The degree of oblateness depends on several factors, namely, the stellar mass, (mean) radius, and the projected rotational velocity \citep{abs08}.}, similar brightness compared to the science star at the wavelength of observation (within $\sim 1$~magnitude), closer than eight degrees on the sky from science target, and, most importantly, to be unresolved point-like sources based on their estimated angular size \citep{van05, boy13a}. Our calibrators, listed in Table~\ref{tab:observations}, have estimated angular diameters ranging from $\theta_{\rm est} = 0.11 - 0.19$~mas \citep{bon06,bon11}.  Our choice of using more than one calibrator with each science star allows the calibrators to be calibrated against one another.  This is important especially when pushing the resolution limits to ensure no unwanted bias is present in the data. 

\begin{figure*}
  \begin{center}
    \begin{tabular}{cc}
      \includegraphics[angle=0,width=8.2cm]{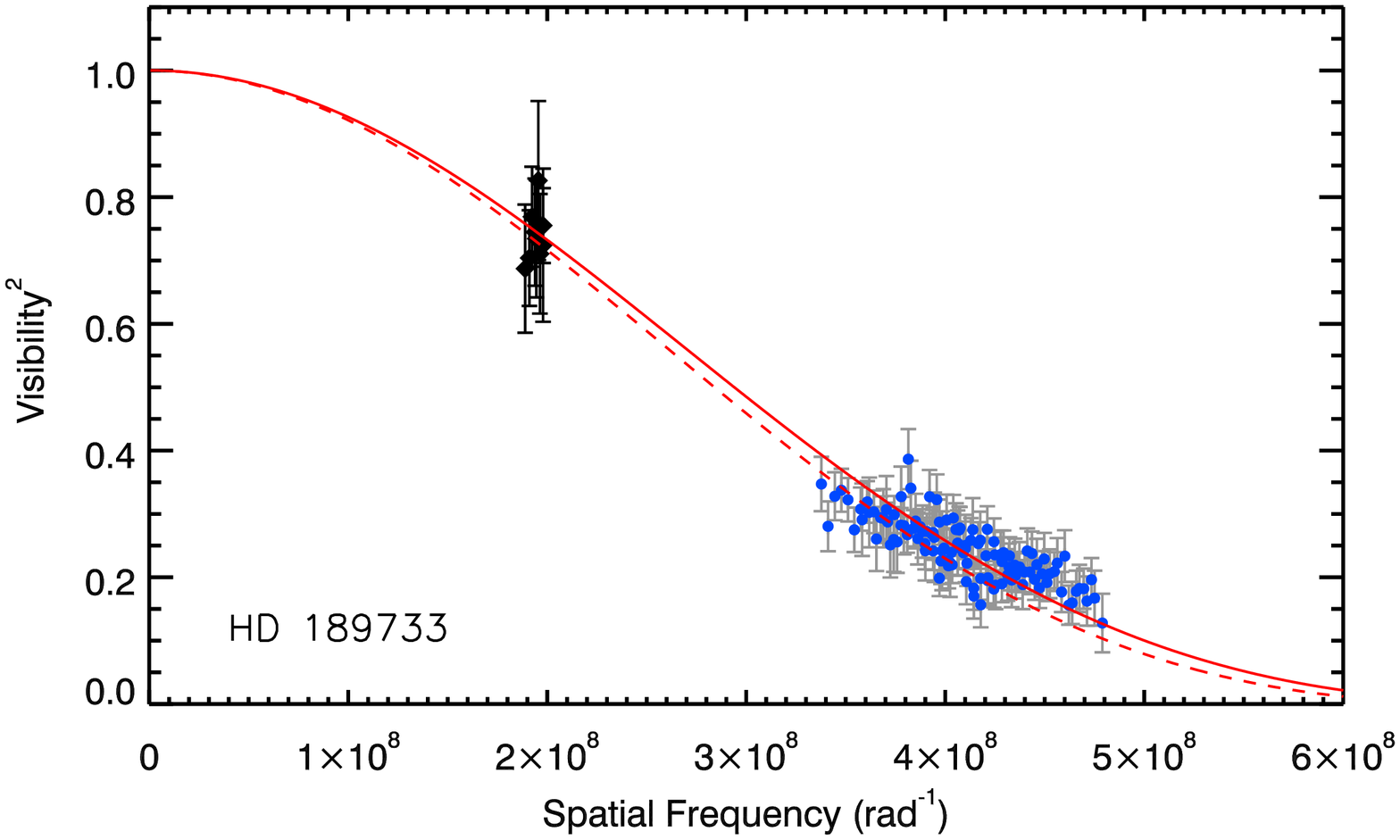}	& 
      \includegraphics[angle=0,width=8.2cm]{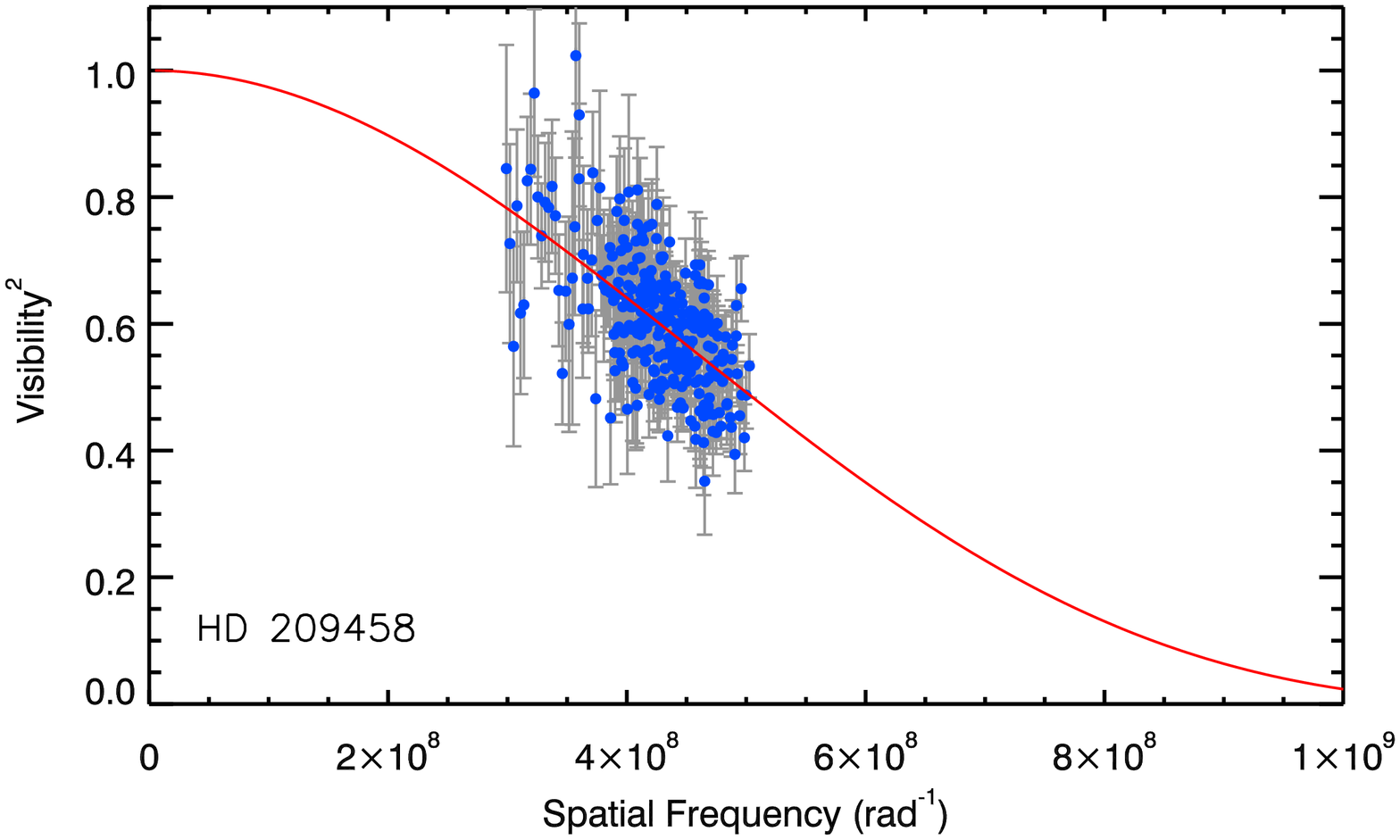} 
     \end{tabular}
  \end{center}
  \caption{Plots of calibrated interferometric $V^2$ values and the limb-darkened $V^2$ values for \one~(left) and \two~(right).  The blue dots are data presented in this work and the solid red line is the $R$-band limb-darkened diameter fit for each star.  The black diamonds are the data from \citet{bai07a}, and the dashed red line represents their $H$-band limb-darkened diameter fit.  Note that the $V^2$ functions are different for \one~due to the limb-darkening coefficient, which is larger in $R$-band compared to $H$-band. The interferometric observations are described in Section~\ref{sec:observations}.}
  \label{fig:visibilities1}
\end{figure*}

Data for each star are reduced and calibrated using the standard reduction routines to extract calibrated squared-visibility measurements ($V^2$) \citep[for details, see][]{mae13,whi13}.  We fit the data to the functions for uniform disk and limb-darkened angular diameters defined in \citet{han74} using the solar metallicity (Table~\ref{tab:properties}), $R$-band linear limb-darkening coefficients from \citet{cla11}.  Limb-darkening is dependent on both the stellar atmospheric properties of temperature and gravity, and we thus iterate on the coefficients to be consistent with the derived stellar properties (see Section~\ref{sec:properties}, Table~\ref{tab:properties}).  Only one iteration was required for the values to converge. The final limb-darkening coefficients we use are $\mu_{R} = 0.67$ and $0.55$ for \one~and \two, respectively. We assume a conservative 5\% uncertainty in these limb-darkening coefficients.  Errors on the fitted angular diameter are computed from a MCMC simulation using 6400 realizations to account for uncertainties in the $V^2$ measurement, in the calibrator diameter (10\%), in limb-darkening coefficients (5\%), as well as the PAVO wavelength scale (5\%) (detailed descriptions are found within \citealt{mae13,whi13}).  We obtain measured uniform disk diameters of $\theta_{\rm UD} =$~\thetaoneud~and \thetatwoud~mas and limb-darkened diameters of $\theta_{\rm LD} =$~\thetaone~and \thetatwo~mas for \one~and \two, respectively. Figure~\ref{fig:visibilities1} shows the data and the visibility curves for each star.  These direct angular diameter measurements agree very well with both stars' predicted angular size using empirically calibrated surface-brightness (SB) relations from \citet{boy14a}: $\theta_{\rm SB} = 0.380 \pm 0.019$ for \one~and $\theta_{\rm SB} = 0.228 \pm 0.011$ for \two, consistent with our measured values to $0.005$ and $0.003$~mas ($0.258$ and $0.002 \sigma$) for \one~and \two, respectively. 

The CHARA Array was also used to measure the diameter of \one~in \citet{bai07a}.  This measurement was obtained using the CHARA Classic beam combiner in $H$-band ($\lambda = 1.67 \mu$m) and yielded an angular diameter of $\theta_{\rm LD} = 0.377 \pm 0.024$~mas (6.7\% error).  Our result presented here for \one~agrees very well (0.008~mas; 0.46~$\sigma$) with this result but reduces the measurement error by a factor of three. The increased precision of our result is due to the choice of beam combiner that operates at shorter wavelengths (samples higher spatial frequencies), which increases the resolution by about a factor of 2.5 times for a given baseline.  To illustrate this difference, we show the data from \citet{bai07a} plotted with our own in Figure~\ref{fig:visibilities1}. 

\citet{bak06b} report the detection of an M-dwarf companion to \one, with separation of $\sim 11$~arcsec.  \citet{bai07a} discussed possible contamination of the interferometric measurements due to this companion, and rejected the possibility.  We confirm that the interferometer's field of $\sim 2$~arcseconds ($\sim 1$~arcsecond mask hole size plus seeing; \citealt{ire08,boy08}) is too little in comparison to the binary separation and thus can not bias the measurements presented here.  

We caution that the angular size of \two\ is at the resolution limit of CHARA/PAVO, and that due to sensitivity limits, its calibrators are at most $\sim 30$\% smaller than our target. Consequently, the measured diameter for \two\ (and subsequently derived stellar properties) may be affected by systematic errors in the estimated calibrator sizes.

\begin{table}
\centering
\caption{Log of interferometric observations\label{tab:observations}}
\begin{tabular}{@{}rccc@{}}
\hline
\textbf{Star} &
 &
\# of &
 \\
 UT Date &
 Baseline &
 Obs  &
 Calibrators \\
\hline
\\

\textbf{HD~189733} \\
2012/05/13 & W1/E1 & 5 & HD~189944, HD~190993 \\ 
2012/05/14 & W1/E1 & 1 & HD~189944, HD~190993 \\ \\

\textbf{HD~209458} \\
2012/08/23 & S1/E1 & 6 & HD~210516, HD~209380, HD~211733 \\ 
2012/08/24 & E1/W1  & 2 & HD~210516, HD~209380, HD~211733  \\ 
2012/10/04 & S1/E1  & 4 & HD~210516, HD~209380  \\ 
2012/11/14 & S2/E2  & 2 & HD~210516, HD~209380  \\  \\
\hline
\end{tabular}
\vspace{-12pt}
\tablecomments{For details on the interferometric observations, see \S\ref{sec:observations}.} 
\end{table}
%

\subsection{Spectroscopic observations} \label{sec:spec}
Optical spectra of \one{} and \two{} were taken with the SuperNova Integral Field Spectrograph \citep[SNIFS,][]{Ald02,Lan04} on the University of Hawaii 2.2m telescope atop Mauna Kea on September 4, 2014. SNIFS split the light into blue (0.32--0.52\um) and red (0.51--0.97\um) channels using a dichroic mirror. The spectral resolution was $\simeq$800 and $\simeq1000$ for the blue and red channels, respectively. Integration times were 33s and 50s for HD 209458 and HD 189733, which yielded a median SNR of $>300$ per resolving element for both stars in the red channel but kept counts below the non-linear region of the detector. 

Details of the SNIFS pipeline can be found in \citet{Bac01} and \citet{Ald06}. Briefly, the pipeline performed dark, bias, and flat-field corrections and cleaned the data of bad pixels and cosmic rays, then calibrated the data based on arc lamp exposures taken at the same telescope pointing and time as the science data. The SNIFS pipeline applied an approximate flux calibration (based on archive data) and collapsed the three-dimensional data cubes into a one-dimensional spectrum using a analytic PSF model. To achieve a more accurate flux calibration and correct telluric lines we used spectra of the EG131, Fiege110, and BD+174708 spectrophotometric standards \citep{Oke90} taken throughout the night and a model of the atmosphere above Mauna Kea \citep{But13}. More details of our SNIFS reduction can be found in \citet{Gai14}.

Near-infrared spectra of \one{} and \two{} were taken with upgraded SpeX \citep[uSpeX][]{Ray03} attached to the NASA Infrared Telescope Facility (IRTF) on Mauna Kea on August 26, 2014. Observations were performed in short cross-dispersed mode with the 1.6$\times15\arcsec$ slit. In this mode uSpeX provided continuous coverage from 0.7\um\ to 2.5\um\ at a resolution of $\simeq400$. Each target was placed at two positions along the slit (A and B) and observed in an ABBA pattern to accurately subtract the sky background by differencing. At least 8 spectra were taken this way, which gave a median SNR$>$200 for each star. To remove effects from large telescope slews, we obtained flat-field and argon lamp calibration sequences after each target. To correct for telluric lines, we observed an A0V-type star immediate after each target and within 0.1~airmasses.

Spectra were extracted using version 4.0 of the SpeXTool package \citep{Cus04}, which performed flat-field correction, wavelength calibration, sky subtraction, and extraction of the one-dimensional spectrum. Multiple exposures were combined using the IDL routine {\it xcombxpec}. A telluric correction spectrum was constructed from each A0V star and applied to the relevant spectrum using the {\it xtellcor} package \citep{Vac03}. The uSpeX orders were merged using the {\it xmergeorders} tool.

Optical and NIR spectra were joined for each star using the overlapping region (0.7-0.9\um), first by scaling the optical to match the NIR data, then by replacing the overlapping region with the weighted mean of the two spectra at each wavelength element. The final spectra reflect the spectral energy distributions (SEDs) for each star with continuous wavelength coverage from 0.32\um\ to 2.5\um. Based on repeated observations taken in the same way and comparisons of spectra from other instruments suggests the relative flux calibration of these spectra are good to better than 1\% \citep{man13}.

\subsection{Bolometric fluxes} \label{sec:fbol}

We calculated bolometric flux ($F_{\rm bol}$) following the procedure from \citet{man13}. To summarize, we obtained flux calibrated literature photometry for each star, which are listed in Table~\ref{tab:SED_phot}. We then computed corresponding synthetic magnitudes from the spectra of each star (Section~\ref{sec:spec}). Each spectrum was scaled to minimize the difference (in standard deviation) between synthetic and literature photometry. 

While zero-points for most of the photometry are generally only calibrated to 1-2\% \citep{boh14}, we use updated zero-points and filter profiles from \citet{Bes12} and Mann \& von Braun (PASP, submitted), which are calibrated to STIS spectra and generally accurate to 1\%. \one\ is known to be variable by 0.03~magnitudes in $V$, although less so at red wavelengths. Both issues are factored in our estimate of the error in $F_{\rm bol}$. Interstellar extinction is set to zero for both targets, due to the small distances to the stars and the unusually tenuous ISM around the solar neighborhood out to a radius of 70~pc \citep{aum09}. This is consistent with the $E(B-V)=0$ for both stars \citep{ram05, arn10}.

\begin{figure*}
  \begin{center}
    \begin{tabular}{cc}
      \includegraphics[angle=0,width=8.2cm]{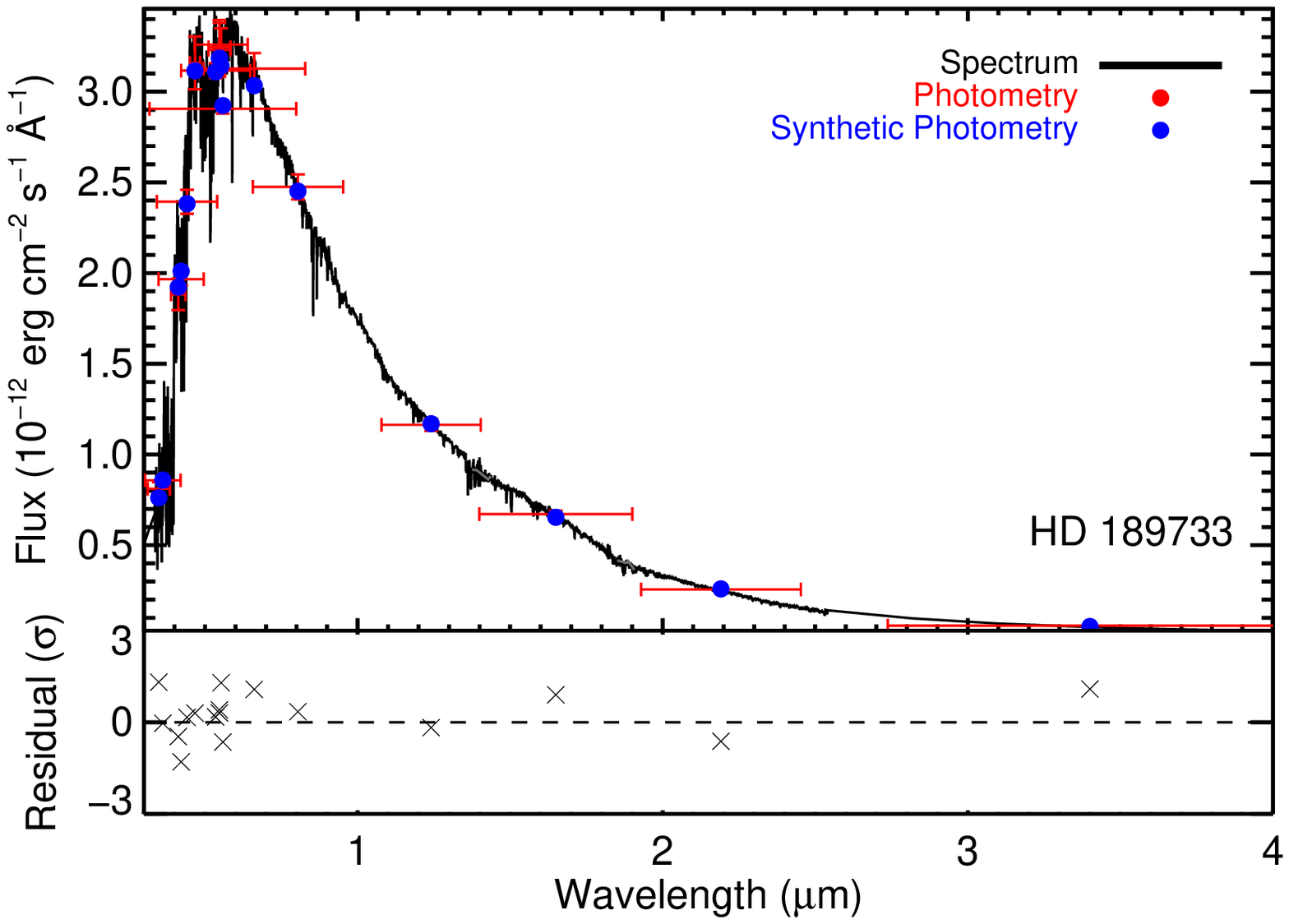}	& 
      \includegraphics[angle=0,width=8.2cm]{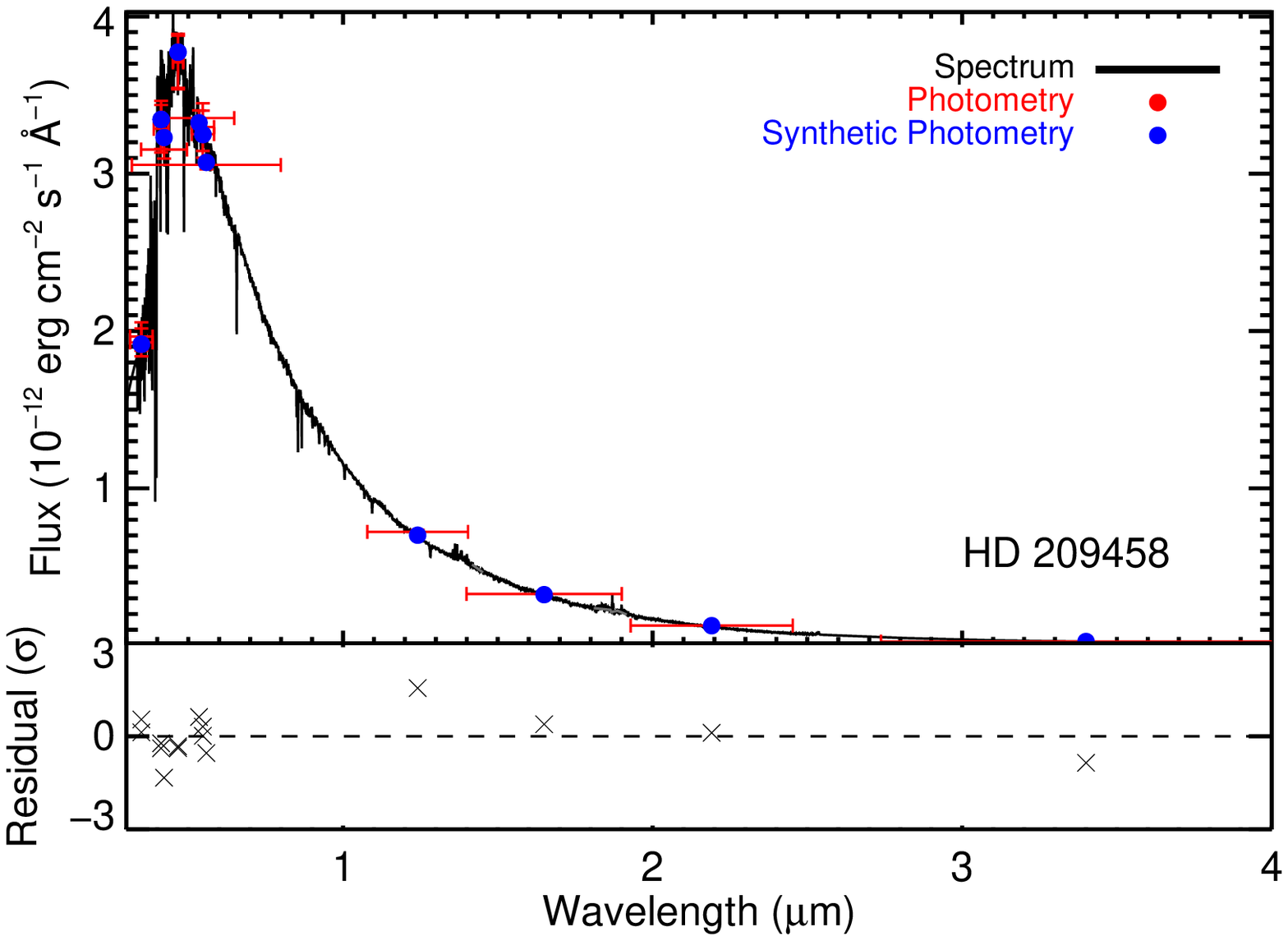} 
     \end{tabular}
  \end{center}
  \caption{SED's for \one~(left) and \two~(right). The (black) spectra represent the joined SNIFS+uSpeX spectra. The (red) points indicate photometry values from the literature. ``Error bars'' in x-direction represent bandwidths of the filters used. The (blue) points show the flux value of the spectral template integrated over the filter transmission. The lower panel shows the residuals around the fit in units of standard deviations. For details, see Section~\ref{sec:fbol}, and for results, see Table \ref{tab:properties}.}
  \label{fig:fbol}
\end{figure*}

We find $F_{\rm bol} = $~\fboloneorig\ and \fboltwoorig\ ($10^{-8}$~erg~s$^{-1}$~cm$^{-2}$) for \one\ and \two, respectively. These bolometric fluxes agree within a percent with values derived in \citet{cas11} via the infrared flux method ($2.7666$\ and $2.3379$, same units). Finally, in order to account for unknown systematic effects due to, for example, uncertainties in photometric magnitude zero point calculations, correlated errors in the photometry, potential errors in the spectral templates, filter transmission functions, etc., we add a 2\% uncertainty to each $F_{\rm bol}$ uncertainty value in quadrature (e.g., see discussion in \citealt{boh14}, in particular their sections 3.2.1 -- 3.2.3). Final results are in Table~\ref{tab:properties}, and we show our calibrated spectra in Figure \ref{fig:fbol}.

\begin{table*}
\centering
\caption{Photometry used in SED fitting}
\label{tab:SED_phot}
\begin{tabular}{lccccl}
\hline
Star & Photometric System  & Filter  & Value &  Uncertainty &  Reference \\	
\hline
HD~189733 	& 	Stromgren 	& 	u 	& 	10.413 & 	0.08 & \citet{1993AandAS..102...89O} \\
HD~189733 	&	Stromgren 	&	v 	&	9.172 	&	0.08  &	\citet{1993AandAS..102...89O} \\
HD~189733 	&	Stromgren 	&	b 	&	8.203 	&	0.08  &	\citet{1993AandAS..102...89O} \\
HD~189733 	&	Stromgren 	&	y 	&	7.676 	&	0.08  &	\citet{1993AandAS..102...89O} \\
HD~189733 	&	Stromgren	&	u 	&	10.4 	&	0.05 &	\citet{2002MNRAS.336..879K} \\
HD~189733 	&	Stromgren	&	b 	&	8.192 	&	0.05 &	\citet{2002MNRAS.336..879K} \\
HD~189733 	&	Stromgren 	&	v 	&	9.161 	&	0.05 &	\citet{2002MNRAS.336..879K} \\
HD~189733 	&	Stromgren 	&	y 	&	7.665 	&	0.05 &	\citet{2002MNRAS.336..879K} \\
HD~189733 	&	Stromgren 	&	y 	&	7.67 	&	0.05 &	\citet{2002MNRAS.336..879K} \\			
HD~189733		&	{\it 2MASS} 		&	J 	&	6.073 	&	0.027  & \citet{cut03} \\
HD~189733 	&	{\it 2MASS} 		&	H &	5.587 	&	0.027  &  \citet{cut03} \\
HD~189733 	&	{\it 2MASS} 		&	Ks	&	5.541 	& 0.015  & \citet{cut03} \\
HD~189733 	&	Johnson	&	U	&	9.241	& 0.1	& \citet{2010MNRAS.403.1949K} \\
HD~189733 	&	Johnson	&	B	&	8.578	& 0.03	& \citet{2010MNRAS.403.1949K} \\
HD~189733 	&	Johnson	&	V	&	7.648	& 0.03	& \citet{2010MNRAS.403.1949K} \\
HD~189733 	&	Cousins	&	Rc	&	7.126	& 0.03	& \citet{2010MNRAS.403.1949K} \\
HD~189733 	&	Cousins	&	Ic	&	6.680	& 0.03	& \citet{2010MNRAS.403.1949K} \\				
HD~189733 	&	Johnson	&	V	&	7.680	& 0.05	& \citet{2011MNRAS.411..435B} \\
\hline
HD~209458 	&	Stromgren 	&	u 	&	9.462 	&	0.08  	&	\citet{1983AandAS...54...55O} \\
HD~209458 	&	Stromgren 	&	v 	&	8.558 	&	0.08  	&	\citet{1983AandAS...54...55O} \\
HD~209458 	&	Stromgren 	&	b 	&	8.020 	&	0.08  	&	\citet{1983AandAS...54...55O} \\
HD~209458 	&	Stromgren 	&	y 	&	7.650 	&	0.08  	&	\citet{1983AandAS...54...55O} \\
HD~209458 	&	Stromgren 	&	u 	&	9.46 	&	0.05 &	\citet{1994AandAS..106..257O} \\
HD~209458 	&	Stromgren 	&	u 	&	9.439 	&	0.05 &	\citet{1994AandAS..106..257O} \\
HD~209458 	&	Stromgren 	&	b 	&	8.018 	&	0.05 &	\citet{1994AandAS..106..257O} \\
HD~209458 	&	Stromgren 	&	b 	&	8.015 	&	0.05 &	\citet{1994AandAS..106..257O} \\
HD~209458 	&	Stromgren 	&	v 	&	8.556 	&	0.05 &	\citet{1994AandAS..106..257O} \\
HD~209458 	&	Stromgren 	&	v 	&	8.548 	&	0.05 &	\citet{1994AandAS..106..257O} \\
HD~209458 	&	Stromgren 	&	y 	&	7.648 	&	0.05 &	\citet{1994AandAS..106..257O} \\
HD~209458 	&	Stromgren 	&	y 	&	7.663 	&	0.05 &	\citet{1994AandAS..106..257O} \\
HD~209458 	&	Stromgren 	&	u 	&	9.443 	&	0.08  	&	\citet{hau98} \\
HD~209458 	&	Stromgren 	&	v 	&	8.546 	&	0.08  	&	\citet{hau98} \\
HD~209458 	&	Stromgren 	&	b 	&	8.011 	&	0.08  	&	\citet{hau98} \\
HD~209458 	&	Stromgren 	&	y 	&	7.650 	&	0.08  	&	\citet{hau98} \\
HD~209458 	&	Johnson 	&	V 	&	7.65 	&	0.01 &	\citet{2000AandA...355L..27H} \\
HD~209458 	&	Johnson 	&	B 	&	8.18 	&	0.02  &	\citet{2000AandA...355L..27H} \\
HD~209458 	&	Johnson 	&	V 	&	7.639 	&	0.02 &	\citet{2001KFNT...17..409K} \\
HD~209458 	&	Johnson 	&	B 	&	8.230 	&	0.04 &	\citet{2001KFNT...17..409K} \\
HD~209458		&	{\it 2MASS} 		&	J 	&	6.591	&	0.011  & \citet{cut03} \\
HD~209458 	&	{\it 2MASS} 		&	H &	6.366 	&	0.035  &  \citet{cut03} \\
HD~209458 	&	{\it 2MASS} 		&	Ks	&	6.308 	& 0.021  & \citet{cut03} \\
HD~209458 	&	Johnson 	&	V 	&	7.693 	&	0.063 &	\citet{2006PASP..118.1666D} \\
HD~209458 	&	Johnson 	&	V 	&	7.640 	&	0.014 &	\citet{2007AN....328..889K} \\
\hline
\end{tabular}
\vspace{-12pt}

\tablecomments{Photometry data used for SED fitting. See \S~\ref{sec:fbol} for details.}											

\end{table*}

%


\section{Stellar and planetary properties}\label{sec:properties}

\subsection{General stellar properties}\label{sec:general_prop}

Hipparcos parallaxes from \citet{van07} are used in combination with our measured angular sizes (Section~\ref{sec:observations}) to determine linear radii for each star.  Furthermore, we are able to calculate the absolute bolometric luminosity for each star with Hipparcos parallaxes and bolometric fluxes from Section~\ref{sec:fbol}.  Lastly, by rearranging the Stefan-Boltzmann equation in terms of observable quantities, we derive effective temperatures for both stars:

\begin{eqnarray}
T_{\rm eff} = 2341 (F_{\rm bol}/\theta_{\rm LD}^2)^{0.25}
\end{eqnarray}

\noindent where the constant 2341 is used for convenient units: the bolometric flux $F_{\rm bol}$ in $10^{-8}$~erg~s$^{-1}$~cm$^{-2}$, the limb darkened angular diameter $\theta_{\rm LD}$ in milli-arcseconds, and the effective temperature $T_{\rm eff}$ in Kelvin.  For both \one~and \two, we present these radii, luminosities, and effective temperatures in Table~\ref{tab:properties}.  The errors on each variable are propagated in quadrature for the final parameter error value listed.

\subsection{The unique circumstances and available auxiliary data for HD 189733 and HD 209458}\label{sec:unique_prop}

Both \one~and \two~are known hosts to transiting exoplanets (\citealt{bou05}; \citealt{cha00,hen00}).  The analysis of a transiting exoplanet's photometric light curve directly measures the planet-to-star radius ratio $R_{\rm p}/R_{\ast}$ \citep{sea11,win10}. 
A transit signature in a photometric light curve is typically confirmed to be planetary in nature with follow-up radial velocity observations.  Such follow-up observations detect radial velocity shifts of the host star from the planet's gravitational pull as it orbits.  This measurement provides the system's mass function, or sum of the masses, when the inclination is known, but not individual masses of both components. New detection techniques have recently allowed for the detection of spectral lines originating from the planet itself (see \citealt{dek14} for details and review of the field).  These measurements of the planet's orbital velocity $K_2$ yields the system mass ratio, $K_2/K_1 = M_1/M_2$, where $K_{1,2}$ are the radial velocity semi-amplitudes and $M_{1,2}$ are the masses of each component.  Thus applying this method to observed transiting planetary systems where the orbital inclination is known from the light curve solution provides {\it absolute} masses of both the host star and the transiting planet, just like an eclipsing binary system.  Currently, our targets are the only two transiting exoplanet systems have been observed in this way. The pioneering work by \citet{sne10} was the first to observe this in \two. \citet{dek13} later announced the successful detection of the planet's radial velocities to the \one~system, which was confirmed with independent efforts in \citet{rod13}.  


In this work, we take advantage of the wealth of knowledge for both the \one~and \two~systems, as described in the above text.  For the remainder of this paper, we assume the $R_{\rm p}/R_{\ast}$ measured from $8 \mu$m Spitzer observations, where the data are least influenced by limb-darkening (references used are \citealt{ago10} for \one~and \citealt{bea10} for \two).  We further use the measured stellar and planetary masses from \citet{dek13} for \one~and \citet{sne10} for \two.  All values and references mentioned are also shown in Table~\ref{tab:properties}.

Using the planet-to-star radius ratio in combination with our measured stellar radius, we are able to {\it empirically} determine the planetary radii of \radonep~\rjup~(2.2\%) and \radtwop~\rjup~(5.4\%) for \one~and \two, respectively. 
Furthermore, knowing both the stellar and planetary mass and radius, it is then straightforward to calculate the surface gravity $\log g$\ ($\log g \propto M/R^2$) and mean density $\rho$\ ($\rho \propto M/R^3$) of each component in the system.  
These values are listed in Table~\ref{tab:properties}. 
The density of \one b~($\rho_{\rm p} =$\ \rhoonecgsp) and \two b~($\rho_{\rm p} =$\ \rhotwocgsp) are much like that of butter and cork, respectively\footnote{http://www.iem-inc.com/information/tools/densities -- ``I can't believe it's not butter'', {\it Fabio}}.

\begin{table*}
\centering
\caption{Stellar and planetary properties}
\label{tab:properties}
\begin{tabular}{@{}lcccc@{}}
\hline
	&	\multicolumn{2}{c}{\one}	&	\multicolumn{2}{c}{\two}	\\
Property &	Value &	Reference &	Value &	Reference \\
\hline
$\theta_{\rm LD}$ (mas)	&	\thetaone	&	this work (\S~\ref{sec:observations})	&	\thetatwo	&	this work (\S~\ref{sec:observations})	\\
$F_{\rm Bol}$	($10^{-8}$~erg~s$^{-1}$~cm$^{-2}$)	&	\fbolone	& this work (\S~\ref{sec:fbol})	&	\fboltwo	&	this work (\S~\ref{sec:fbol})	\\
$L_{\ast}$ (L$_{\odot}$)	&	\luminone	&	this work (\S~\ref{sec:general_prop})	&	\lumintwo	&	this work (\S~\ref{sec:general_prop})	\\
$R_{\ast}$ (R$_{\odot}$)	&	\radone	&	this work (\S~\ref{sec:general_prop})	&	\radtwo	&	this work (\S~\ref{sec:general_prop})	\\
$T_{\rm eff}$ (K) 	&	\teffone \phn\phn	&	this work (\S~\ref{sec:general_prop})	&	\tefftwo \phn\phn	&	this work (\S~\ref{sec:general_prop})	\\	
$[$Fe/H$]$	(dex)		&	$-0.03 \pm 0.08$\phs	&	\citet{tor08}	&	$0.00 \pm 0.05$	&	\citet{tor08}	\\
	$R_{\rm p}/R_{\ast}$	&	$0.155313\pm0.000188$	&	\citet{ago10}	&	$0.12403\pm0.00043$	&	\citet{bea10}	\\
$R_{\rm p}$ (R$_{\rm Jup}$)	&	\radonep	&	this work (\S~\ref{sec:unique_prop})	&	\radtwop	&	this work (\S~\ref{sec:unique_prop})	\\
	$M_{\ast}$ (M$_{\odot}$)	&	$0.846\pm0.049$	&	\citet{dek13}	&	$1.00\pm0.22$	&	\citet{sne10}	\\
	$M_{\rm p}$ (M$_{\rm Jup}$)	&	$1.162\pm0.058$	&	\citet{dek13}	&	$0.64\pm0.09$	&	\citet{sne10}	\\
$\log g_{\rm p}$	&	\loggonep	&	this work (\S~\ref{sec:unique_prop})	&	\loggtwop	&	this work (\S~\ref{sec:unique_prop})	\\
$\log g_{\ast}$	&	\loggone	&	this work (\S~\ref{sec:unique_prop})	&	\loggtwo	&	this work (\S~\ref{sec:unique_prop})	\\
$\rho_{\rm p}$ ($\rho_{\rm Jup}$)	&	\rhoonep	&	this work (\S~\ref{sec:unique_prop})	&	\rhotwop	&	this work (\S~\ref{sec:unique_prop})	\\
$\rho_{\ast}$ ($\rho_{\odot}$)	&	\rhoone	&	this work (\S~\ref{sec:unique_prop})	&	\rhotwo	&	this work (\S~\ref{sec:unique_prop})	\\
\hline
\end{tabular}
\end{table*}


\section{Previously determined host star properties} \label{sec:discussion}


The first direct measurement of \one's radius was made by \citet{bai07a} (Section~\ref{sec:observations}).  We have shown that our data, taken at much higher resolution, agree with this result by well under one sigma, as well as improve the error by a factor of $4.5$. No prior direct measurements of the radius of \two~are published for comparison.  

Over the years, estimates of the stellar properties of each star have been made using many techniques. We compare our values to the transiting exoplanet host star properties from \citet{tor08} and \citet{sou10,sou11}\footnote{The planetary density in \citet{sou10} is corrected in \citet{sou11} using the right scaling constant for Jupiter's density, effectively lowering previous densities by $\sim 7$\%.  \citet{sou09} provides a lot of the background framework to the sequentially later papers cited here.}.  The \citet{tor08} and \citet{sou10} papers both consist of a rigorous, uniform analysis using all available literature data on known transiting systems at the time. Similarly, their efforts make use of the photometric ($a/R_{\ast}$) measured from the light curve \citep{sea03} as an external constraint on surface gravity, expanding upon the method developed by \citet{soz07}.  \citet{tor08} derive stellar properties (mass, radius, luminosity, surface gravity, and age) by fitting Yonsei-Yale ($Y^2$) evolutionary models \citep{yi01,yi03,dem04} to the spectroscopically determined $T_{\rm eff}$ and [Fe/H], using the photometric ($a/R_{\ast}$) as evolutionary indicator.
Host star properties derived in \citet{sou10, sou11} are derived by a similar approach, using up to six different evolutionary models as well as empirically established relations derived from well-studied eclipsing binaries. Other select references to determine stellar properties are also touched upon in the discussion to follow, though the vast amount of literature references for each star makes a complete comparison demanding, with very little return.  

The mean stellar density computed by our method (Table~\ref{tab:properties}) and the density determined via the photometric time series analysis are the most fundamentally derived values to compare, since they are largely independent of models.  The stellar density derived for \one~and \two~agree well with our measurements within $\sim 1.5 \sigma$ for \one~and within $1.0 \sigma$ for \two~\citep{sou10,tor08}.

Stellar radii are determined indirectly, generally via stellar evolutionary models, using the results from high-resolution spectroscopic observations with stellar atmosphere models as inputs (see above).  In this way, \citet{tor08} find the radius for \two~to be $1.155 \pm 0.015$~\rsun, agreeing well with our value within $0.8 \sigma$~($0.05$~\rsun). The detailed, yet indirect estimate of \two's stellar radius by \citet{cod02} of $R = 1.18 \pm 0.1$~\rsun\ agrees with our value within $0.2 \sigma$~($0.02$~\rsun).  Note that since \two~is located at a distance of nearly $50$~pc, the errors in the {\it Hipparcos} parallax contribute significantly to our linear radius calculation.  For this work, we assume the parallax from the \citet{van07} reduction ($\pi = 20.15 \pm 0.80$~mas; distance~$= 49.63 \pm 1.97$~pc).  However, if we were to use the \citet{esa97} parallax value from the first {\it Hipparcos} reduction ($\pi = 21.24 \pm 1.00$~mas; distance~$= 47.08 \pm 2.22$~pc), our radius measurement would be $R_{\ast} = 1.14 \pm 0.06$~\rsun. While this radius value is still consistent within errors of the adopted values mentioned above, it underlines the importance of having an accurate distance measurement to \two~in order to constrain our results better.

On the other hand, the radius for \one~from \citet{tor08}, $R_{\ast} =0.756 \pm 0.018$~\rsun, is $2 \sigma$~($0.05$~\rsun) smaller than our measurement.  The significant offset of the \citet{tor08} radius and our measurement for \one~is likely a result from the evolutionary model not being able to reliably reproduce observed stellar parameters in later-type stars (e.g., \citealt{boy12b}).  This detail was addressed in \citet{tor08} for the M-dwarf transiting planet host GJ\,436, and thus the stellar properties for that star came from a specialized method described in \citet{tor07}. This semi-empirically determined radius value of GJ\,436 was confirmed by \citet{von12}, who directly measured its radius using LBOI. The two values agree by $\sim 0.4 \sigma$ (2\%).  The evolutionary model predictions however, yield a radius $>10$\% smaller for this star, a known shortcoming in the models for low-mass stars, as discussed in \citet{tor08} and \citet{von12}.

\begin{figure*}
  \begin{center}
      \includegraphics[angle=0,width=12.2cm]{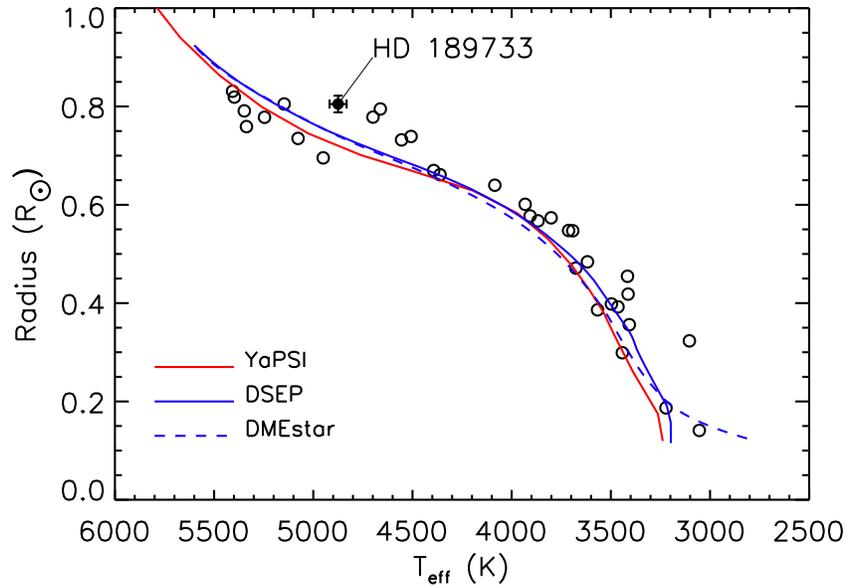}
  \end{center}
    \caption{Radius - temperature plot showing the position of \one~(filled point) with the low-mass stars in~ \citet{boy12b} (open points). Also plotted are solar abundance, 5~Gyr isochrones from the DSEP (solid blue line), DMEstar (dashed blue line), and YaPSI (solid red line) model grids. Refer to Section~\ref{sec:discussion} for details.}
    \label{fig:Radius_VS_Temp_add_HD189733}
  \end{figure*}

While this deficiency in stellar models is generally viewed as a concern for the stellar properties of M-dwarfs, similar incompatibilities exist for more massive stars.  As shown in \citet{boy12b}, the observed radii and temperatures of single, K- and M-dwarfs were discrepant with the predictions from the Dartmouth models (DSEP; \citealt{dot08}).  Specifically, \citet{boy12b} found that models over estimate temperatures by $\sim 3$\%, and under estimate radii by $\sim 5$\% for stars cooler than about $5000$~K.  This discrepancy was independently confirmed by \citet{spa13} using YaPSI (Yale-Potsdam Stellar Isochrones), the most recent set of tracks and isochrones calculated with the Yale Rotational stellar Evolution Code (YREC).  In Figure~\ref{fig:Radius_VS_Temp_add_HD189733}, we show the measured radius and effective temperature of \one~(solid point) with the low-mass stars that have directly measured radii and temperatures in \citet{boy12b} (open points).  

Also displayed in Figure~\ref{fig:Radius_VS_Temp_add_HD189733} are solar metallicity, 5~Gyr isochrones from the Dartmouth Stellar Evolution Database (DSEP; \citealt{dot08}),  DMEstar (Dartmouth Magnetic Evolutionary Stellar Tracks And Relations; updated DSEP grid of models, described briefly in \citealt{mui14} and \citealt{mal14}), as well as YaPSI (\citealt{spa13}).  Figure~\ref{fig:Radius_VS_Temp_add_HD189733} shows that most of the points with $T_{\rm eff} <5000$~K fall above the model isochrone predictions.  The position of \one~in this plot is consistent with the parameter space where model predictions deviate from the directly measured astrophysical properties for the lower-mass stars \citep{spa13, boy12b}.

\section{Harmonizing stellar evolutionary model predictions with observational data}\label{sec:modelcomp}


In Section~\ref{sec:discussion}, we show that the properties of \two~are consistent with model predictions, however, \one~shows potential for significant disagreement.  As an initial comparison to models, the properties of \one~are interpolated onto a 5~Gyr, solar metallicity isochrone from the Dartmouth series \citep{dot08}. When using stellar luminosity as the dependent variable, the model predicted mass is $0.805$~\msun, consistent with observations.  However, the model \teff~$= 5028$~K and $R = 0.756$~\rsun\ are 150~K ($3.6\sigma$)~too hot and 0.05~\rsun\ ($2.9 \sigma$)~too small, respectively. If the empirical mass is used as the dependent variable instead, models predict $R = 0.796$~\rsun, \teff~$= 5225$~K and $L = 0.4245$~\lsun. In this scenario, the radius is consistent with observations, but the \teff~and luminosity are too high by 350~K ($8 \sigma$)~and 0.1~\lsun\ ($9\sigma$), respectively. Model predictions are therefore not compatible with the empirical data. 

Here, we explore various explanations for the discrepancies between the model predictions and the empirical data.  We investigate the effects of age, composition, and how convection is treated in models given the constraints provided by the observational data.  Given the observational results (Table~\ref{tab:properties}), we are able to deduce the likely cause of the model offsets for the predicted properties of \one~is due to the treatment convection and the choice of the mixing-length parameter \amlt\ (see Section~\ref{sec:amlt}).

\subsection{Age}\label{sec:age}
Models with masses in the $0.8$~\msun\ range undergo non-negligible evolution
along the main sequence. Thus, adoption of a 5~Gyr isochrone is not exactly
appropriate. Allowing for variations in age is equivalent to investigating 
whether models of different masses provide a more consistent fit to the data.
However, we find that no models simultaneously fit the \teff\ and radius of \one\ at an age
younger than that of the Universe. In all cases, when a given model mass
track fits the measured radius, the \teff\ is too hot. Conversely, when the
models match the measured \teff, the radius is too small. However, this is
only considering ages along the main sequence.

Along the pre-main-sequence (pre-MS), at an age near 40~Myr, models with masses around
$0.82$~\msun\ match the complete set of observed properties. Around 40~Myr, models
suggest a star of this mass is nearing the main-sequence, having developed a 
small convective core prior to the $p-p$ chain coming into full equilibrium,
which brings about the establishment of a radiative core. Although these pre-MS 
models are in agreement with the observed stellar properties, other factors 
indicate that \one\ is unlikely to be a pre-MS star. Evidence comes 
from its derived rotation period ($\sim 11$~days), its low levels 
of magnetic activity \citep{gui13,pil10}, and lack of any detectable lithium 
in the spectrum \citep{mis12}, all of which indicate \one\ is a MS star. 

\subsection{Composition}\label{sec:composition}
Departures from a strictly-scaled solar composition could, in principle, 
provide better agreement between observations and the stellar models. These 
departures include differences in the bulk metallicity, variations in 
$\alpha$-element abundances, a non-solar helium abundance, and also the solar 
heavy element mixture.

\subsubsection{Metallicity}
To fit the observed \teff\ and radius, models require a scaled solar metallicity
of [M/H]~$= +0.2$~dex and predict an age of approximately 10~Gyr. Abundance analyses,
however, find [M/H]$= -0.03 \pm 0.08$~dex, with a tendency for mildly sub-solar 
metallicity \citep{bou05,tor08}. Individual element abundances show variation consistent with
[M/H] within the uncertainties. One prominent exception is oxygen, which appears 
under-abundant in \one~by roughly 0.2~dex ([O/Fe]~$ = -0.2$~dex; \citealt{mis13}) compared to 
the Sun. Thus, there does not appear to be evidence for a super-solar metallicity 
in \one's atmosphere which is required by the models to fit the observations.

\subsubsection{$\alpha$-element enhancement}
In light of the measured under-abundance of oxygen, it is possible that the star
has a non-solar-like abundance of $\alpha$-elements. We test this idea with 
models having [$\alpha$/Fe]~$= -0.2$ and $+0.2$~dex as a proxy for variations in 
oxygen abundance. Increasing [$\alpha$/Fe] has the effect of reducing both the
\teff\ and radius of the model whereas decrease [$\alpha$/Fe] has the opposite
effect. Agreement is found using [$\alpha$/Fe]~$= +0.2$~dex at an age
of roughly 9~Gyr, but this is in disagreement with the observed oxygen abundance
of \one~(see above). Models incorporating individual element enhancement, 
in particular carbon, nitrogen, and oxygen, are needed to further assess 
whether departures from a strictly solar abundance provide better agreement.

\subsubsection{Helium abundance}
The abundance of helium in the standard models presented thus far is set by assuming
the helium mass fraction scales linearly with bulk metallicity from the primordial 
value $Y_p = 0.2488$~\citep{pei07}. However, variations in the assumed helium 
abundance can have a significant impact on stellar models through changes 
in the mean molecular weight. Reducing the helium abundance effectively 
leads to a lower \teff\ and a smaller radius due to reductions 
in the {\it p--p} chain energy generation rate. 

Dartmouth models were generated with $Y = 0.24$, 0.25, 0.26, and 0.278, where 
the latter value is the solar calibrated value for a model with solar metallicity. Only by reducing
the initial helium abundance of the models below $Y = 0.25$ is it possible to find
a model that reproduces the observed properties of \one. It is worrisome that the
required helium abundances are below the primordial value, leading us to doubt that
helium abundances variations are a plausible explanation. 

\subsubsection{Solar mixture}
Along the same lines as reducing the proportion of $\alpha$-elements and the overall
helium abundance, it is possible that the solar heavy element mixture is incorrect
in the standard models adopted here. Standard Dartmouth models adopt the abundances
from \citet{gre98}, despite trends in the literature toward a lower heavy
element composition (e.g., \citealt{asp09,caf11}). As a test, we 
computed a set of Dartmouth models adopting the \citet{asp09} solar composition
after first re-calibrating the models to the Sun. 

We find that it is possible to reproduce the properties of \one\ with an $0.80$\msun\ model at 
an age of 7~Gyr using the \citet{asp09} abundances. It is encouraging that 
agreement can be found, but caution must be exercised as there are significant unresolved
issues between helioseismic data and standard solar models that adopt the Asplund et al.
abundances (see, e.g., \citealt{bas08,bas13}). Since solar models calculated with 
the Asplund et al. abundances do not provide an adequate representation
of the solar interior, any agreement found with other stars must be regarded with 
skepticism.

\subsection{Convection}\label{sec:convection}

One final aspect of stellar modeling that we wish to address is the efficiency of
thermal convection. This is relevant considering recent results from asteroseismic
studies, suggesting that convective properties are dependent on intrinsic stellar
properties such as mass and composition \citep{bon12} and 
the on-going issue regarding inflated radii of low-mass stars 
in detached eclipsing binaries (e.g., \citealt{tor10}).

\subsubsection{Reduced \amlt}\label{sec:amlt}

A simple test is to compute models with various convective mixing length parameters.  Doing so with the Dartmouth models, we find that a mixing length parameter of \amlt~$= 1.4$ is required to bring an $0.81$~\msun\ model into agreement with the observations. By comparison, the relationship between stellar properties and convective mixing length parameter suggested by \citet{bon12} predicts a mixing length of \amlt~$= 1.44$, when re-scaled to the solar-calibrated mixing length in the Dartmouth models. The close  agreement may imply that the disagreement between models and the observations is  the results of natural variations in convective efficiency. However, we must note that \one, with an empirically determined \teff~$=$~\teffone~K is outside of the calibration range of the \citet{bon12} relation. 

\subsubsection{Making constrained models}\label{sec:yrec}

Using the directly measured stellar properties, we are able to empirically test how \amlt\ will change to find agreement with stellar evolutionary models. Although models for \two\ do not have difficulty reproducing its observables - likely due to its closer similarity to the Sun - we apply this test on both stars studied here.  We use the observed radius, temperature, mass and associated errors to generate YREC models in a Monte Carlo analysis.   
In this mode, models are constructed to satisfy the observed mass, radius, temperature and metallicity constraint. The 
mixing length parameter is varied and age (and initial helium abundance) is a free parameter. For each run,
mass, radius, effective temperature and metallicity are varied assuming that their errors have a Gaussian distribution.
This requires the code to run in an iterative mode. In all cases only models with ages $< 13.8$~Gyr
are chosen. We use standard physics inputs for the models. We use the OPAL equation of state
\citep{rogersandnayfonov}. We use high temperature opacities from OPAL \citep{iglesiasandrogers} and supplemented them
with low temperature opacities from \citet{lowtemp}. We use nuclear reaction rates of \citet{adelberger} except for the
$^{14}$N($p$,$\gamma$)$^{15}$O reaction, where we use the reaction rate of \citet{formicola}. Gravitational
settling and diffusion of helium and heavy elements are incorporated using the coefficients of
\citet{thoul}.

\begin{figure*}
  \begin{center}
    \begin{tabular}{cc}
      \includegraphics[angle=0,width=8.2cm]{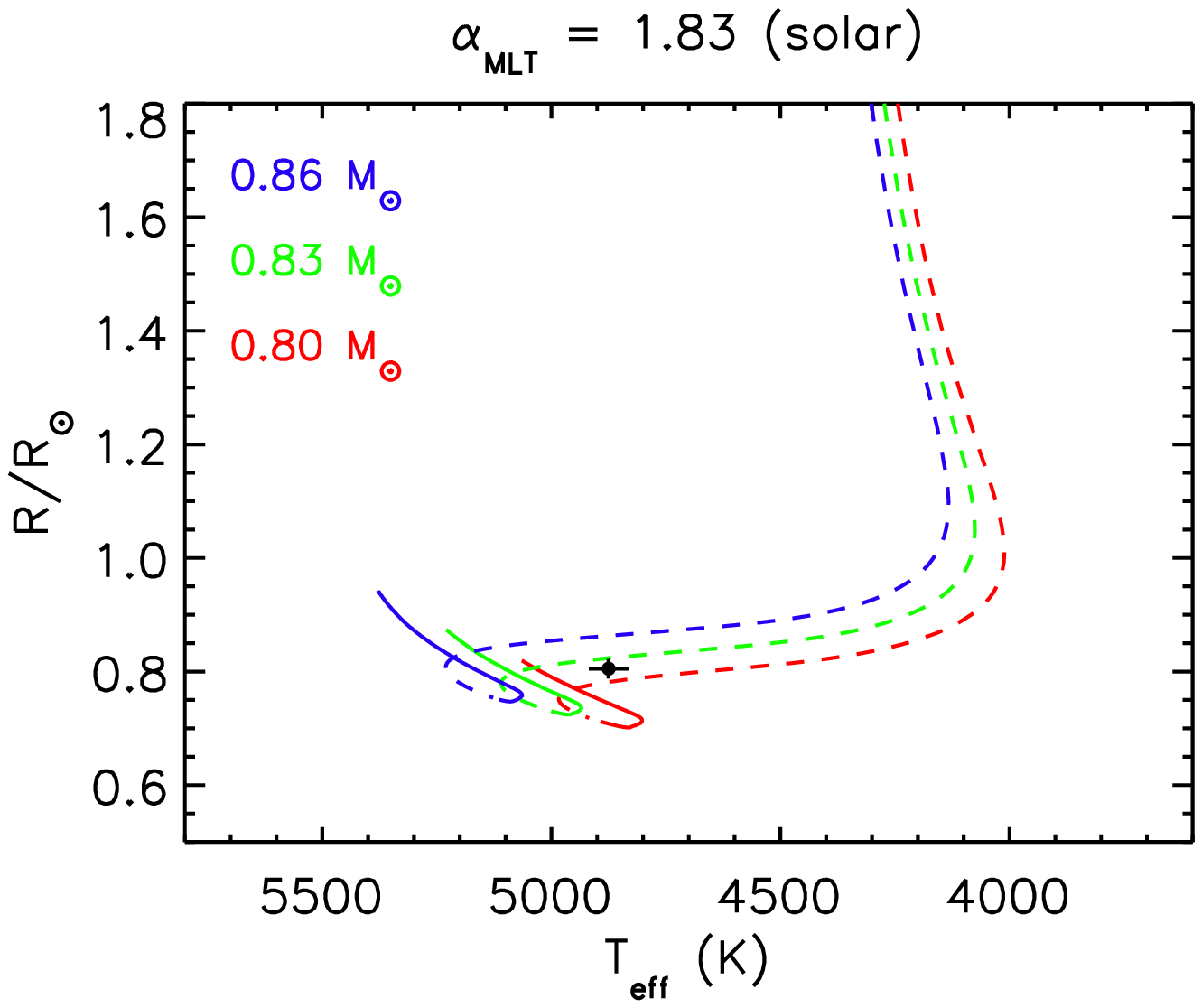}	& 
      \includegraphics[angle=0,width=8.2cm]{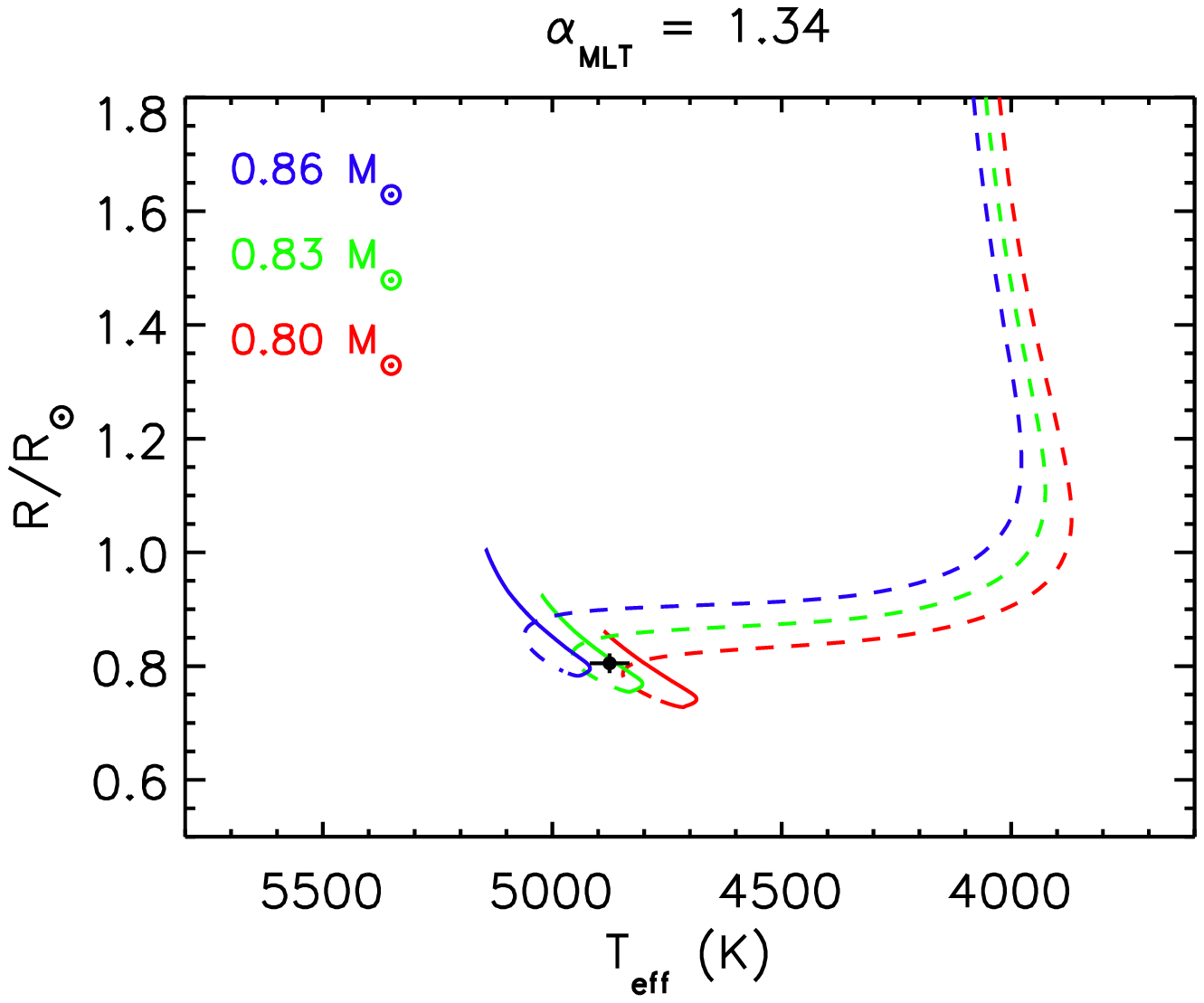} 
     \end{tabular}
  \end{center}
  \caption{Radius-temperature plot showing the position of \one~with 1-$\sigma$~errors (black point).  Each panel also shows evolutionary tracks for the mass indicated in legend, corresponding to the mass of \one~(green), and a range above (blue) and below (red) this value by $\sim 1$-$\sigma$ of the MC analysis. Dashed and solid lines denote pre-main sequence and main sequence evolutionary stages, respectively. The left panel are models using the solar calibrated \amlt, and the right panel is \amlt\ parameter found in this work. Note, only models with a reduced \amlt\ reproduce the observed stellar properties for \one\ as a main-sequence star. For details, see Section~\ref{sec:amlt}~and \ref{sec:yrec}.}
  \label{fig:amlt}
\end{figure*}

Since many of the (mass, radius, $T_{\rm eff}$, $Z$) combinations for a given mixing length parameter
end up requiring an initial helium abundance less than that produced by the Big Bang, the results of the
Monte Carlo are analyzed in two ways.  The first way, \textquotedblleft unconstrained\textquotedblright, accepts all results. The other way, \textquotedblleft constrained\textquotedblright, only allowed results for which the 
initial helium abundance was greater than the primordial value $Y_p = 0.2488$ \citep{pei07}. The results 
of the analysis are shown in Table~\ref{tab:model_outputs} and in Figure~\ref{fig:amlt}, where the \textquotedblleft constrained\textquotedblright~solution for \one~yields \amlt~$= 1.34 \pm 0.18$, a significantly lower \amlt\ compared to a solar-mass star.  This result illustrates that with standard physics, a change in the mixing length parameter is enough to obtain physical models of the two stars, \one~and \two.


\begin{table*}
\centering
\caption{YREC model outputs\label{tab:model_outputs}}
\begin{tabular}{@{}rcccc@{}}
\hline
&   \multicolumn{2}{c}{\textbf{\one}} &  \multicolumn{2}{c}{\textbf{\two}}  \\
\textbf{Property} & 
 Unconstrained &
 Constrained\tablenotemark{a} &
 Unconstrained &
 Constrained\tablenotemark{a} \\
\hline

\amlt 					&	$1.65 \pm 0.38$		&	$1.34 \pm 0.18$		&	$2.01 \pm 0.43$		&	$2.01 \pm 0.43$		\\ 
Age	(Gyr)				&	$5.2 \pm 3.5$		&	$4.3 \pm 2.8$		&	$6.5 \pm 2.7$		&	$6.5 \pm 2.7$ 		\\
Initial helium (Y$_0$)	&	$0.228 \pm 0.031$	&	$0.266 \pm 0.016$	&	$0.3240 \pm 0.0377$	&	$0.3241 \pm 0.037$	\\
Helium (Y)				&	$0.216\pm 0.032$	&	$0.252\pm 0.016$	&	$0.2808 \pm 0.0388$	&	$0.2809 \pm 0.0387$	\\	
\hline
\end{tabular}
\tablenotetext{a}{With initial helium Y$_0 > 0.2488$}
\tablecomments{See \S~\ref{sec:yrec} for additional details.} 
\end{table*}



\subsubsection{Magneto-convection}
As an alternative explanation for the reduced convective mixing length, we computed
magnetic stellar evolution models using the Dartmouth code DMEstar \citep{fei12b,fei13}. 
Two approaches to modeling the influence of the magnetic field were adopted:
one that mimics a rotational dynamo by stabilizing convective flows and another whereby
convective efficiency is reduced so as to mimic a turbulent dynamo. The magnetic models
(of both varieties)
require surface magnetic field strengths of approximately 1.5~kG to reproduce the 
observations. \citet{fei13} showed that, to a reasonable extent, the interior
magnetic field is of less consequence than the surface magnetic field in stars with a 
radiative core. Thus, the requirement of a 1.5~kG magnetic field is fairly robust, 
unless super-MG magnetic fields are invoked in the interior. While \one~is fairly
active in comparison to the Sun, the measured magnetic field is constrained to be 
in the range of 40 -- 100~G \citep{mou07,pil14}, considerably lower than required by the models.

\subsubsection{Star spots}
Finally, \one~is known to show light curve modulations consistent with the
presence of spots on the stellar surface. On short timescales, spots reduce the 
flux leaving the stellar surface without influencing the star's radius 
(e.g., \citealt{spr86}). This would lower the observed luminosity and \teff, 
producing disagreement between models and observations. If we assume that stellar
evolution models reproduce the correct radius, but over-estimate the \teff, 
then one can estimate the potential spot coverage required to produce the 
observed luminosity difference. 

At the observed radius, standard evolutionary models 
predict \one~to have a mass consistent with observations of approximately
$0.85$~\msun, but a luminosity 35\% higher than observed. Following \citet{cha07}, 
this luminosity difference implies spot coverages of between 51\% -- 73\%
if the spots are 25\% -- 15\% cooler than the surrounding photosphere, respectively.
At this level, spots should be detectable using either Dopper Imaging or by
modeling spectral features (e.g., \citealt{one98}). In fact, this level of 
spottiness is not consistent with observations of molecular features for 
even the most active stars \citep{one98}.  Alternatively, a significant
coverage of spots would produce anomalous photometric colors compared to predictions
from non-spotted stellar models. However, an $0.85$~\msun\ stellar model from the 
Dartmouth series predicts the correct photometric magnitudes and colors. Introducing
deviations due to spots produces worse photometric agreement, and is thus unlikely 
the cause of the offset.

\section{Summary}\label{sec:conclusion}

We present direct measurements to the physical properties of two Hall of Fame transiting exoplanet host stars, \one~and \two. We use the CHARA Array to measure the stellar angular diameters. By combining these measurements with distance and bolometric flux, we determine the linear radius, effective temperature, and absolute luminosity for each star (Table~\ref{tab:properties}). Combined with the empirically determined dynamical masses \citep{dek13,sne10}, and the planet-to-star radius ratio from Spitzer data \citep{ago10,bea10}, we are able to calculate full system properties for both star and planet independent of models.

We find that the observations of \two\ agree with evolutionary model predictions.  However, the properties of \one\ show discrepancies with models not unlike previously seen with fundamental measurements of low-mass stars \citep{boy12b}.  We consider several scenarios in the attempt to reconcile the differences in either the assumed stellar properties or standard input physics within the models.  We conclude that the models will match the data only by adjusting the solar-calibrated mixing length parameter to a lower value (Section~\ref{sec:amlt}). This work highlights the importance in calibrating \amlt\ for stars with masses less than the Sun.  As such, if models remain unchanged, the trend of models predicting temperatures too high and radii too small will remain.  This has significant impact on the field of exoplanet detection and characterization, particularly in the case for low-mass stars too small/faint to be resolved with LBOI \citep{man13}.

The analysis and discussions within this work primarily focus on the discrepancy between our observations and evolutionary model predictions.  As such, we do not address in detail comparisons with stellar properties derived with high-resolution spectroscopy, which are heavily model dependent and have sparse empirical verification.  However, it is worthy to note that the temperature estimates listed in the PASTEL Catalogue of stellar parameters \citep{sou10a} for \one\ range from $4952 - 5111$~K, our temperature being 77~K cooler than the lowest entry. The temperature we measure for \two\ falls in the middle of the range in the PASTEL Catalogue ($5987 - 6142$~K). We can only speculate the reason for this large discrepancy in the temperature for \one\ is due to an extra source of opacity, such as TiO, which begins to appear at this temperature, that is not being correctly accounted for in the models.  Another possible reason for the discrepancy is that if the spectroscopic modeling identifies an incorrect $\log g$, this will bias the resulting \teff\ and metallicity estimates \citep{buz01}. Likewise, the semi-empirical approach to determine \teff\ using the Infrared Flux Method (IRFM; \citealt{bla79}) has been refined over the years to incorporate many details with goals to establish a effective temperature scale to better than 1\%.  While the IRFM is a semi-empirical approach, systematics up to 100~K between IRFM scales (e.g., \citealt{gon09,cas10}, and references therein) exist, where the differences may be associated with lack of empirical measurements (i.e., interferometry) to calibrate zero-points \citep{boy13a}.  Particularly for stars with $T_{\rm eff} < 5100$~K, IRFM temperatures are systematically hotter by a few percent \citep[][their figure 20]{boy13a}.  This statement holds true for \one, where the IRFM temperature of $5022$~K from \citet{cas11} is 150~K (3\%) hotter than the interferometric \teff\ derived in this work.  The fact that spectroscopic and IRFM estimates of \one's \teff\ are considerably higher than the interferometric value is further evidence that indirect estimates of cool star properties need to be used with caution until they are able to be calibrated with empirical data sets.

A further implication of the corrections to stellar parameters is the calculated extent of the Habitable Zone \citep{kop13,kop14}. \citet{kan14} quantified the importance of stellar parameter determinations in defining the HZ boundaries for a particular system. Although the known planets in the systems studied in this paper cannot be consider HZ planets, the divergence of the measured stellar parameters from stellar models will have serious consequences for correct determinations of the fraction of stars with Earth-sized planets in the HZ ($\eta_{\oplus}$). This is particular true for late type stars since (i) the short-period bias of the transit and radial velocity methods is preferentially revealing $\eta_{\oplus}$ for this stellar population, and (ii) calculated late-type stellar properties tend to have the largest divergence from models. It is therefore of critical importance to consider these results when describing HZ regions for current and upcoming targets, such as those of the Transiting Exoplanet Survey Satellite (TESS) \citep{ric14}.


\section*{Acknowledgments}

TSB acknowledges support provided through NASA grants ADAP12-0172 and 14-XRP14\_2-0147. DH acknowledges support by NASA Grant NNX14AB92G issued through the Kepler Participating Scientist Program. SB acknowledges partial support of NSF grant AST-1105930. Judit Sturmann keeps some tight beams in place - hats off to you girl!  The CHARA Array is funded by the National Science Foundation through NSF grants AST-0606958 and AST-0908253 and by Georgia State University through the College of Arts and Sciences, as well as the W. M. Keck Foundation. This research made use of the SIMBAD and VIZIER Astronomical Databases, operated at CDS, Strasbourg, France (http://cdsweb.u-strasbg.fr/), and of NASA's Astrophysics Data System, of the Jean-Marie Mariotti Center \texttt{SearchCal} service (http://www.jmmc.fr/searchcal), co-developed by FIZEAU and LAOG/IPAG.




\bibliographystyle{mn2e}            

\bibliography{mn-jour,paper}      


\end{document}